\documentclass[12pt]{book}
\usepackage{plenum}
\usepackage{epsfig}
\usepackage{amssymb}
\begin{document}
\chapter{PRIMORDIAL MAGNETIC FIELDS AND THEIR DEVELOPMENT (APPLIED FIELD
THEORY)}

\author{P. Olesen} 

\affiliation{The Niels Bohr Institute\\
Blegdamsvej 17\\DK-2100 Copenhagen \O\ \\Denmark}

\vskip2cm

{\sl Invited talk at the ``Nato Advanced Research Workshop on Theoretical
Physics'', June 14-20, 1997, Zakopane, Poland}\\

\section{MOTIVATION}

In this talk I discuss the non-linear development of magnetic fields
in the early universe. Since this is based on a classical field theory, which
turns out to have a rather complex structure, I thought it could be
of interest for this meeting, under the heading of ``applied field theory''.
I very briefly mention a number of particle physics mechanisms for generating
magnetic fields in the early universe, but the emphasis is on the
field theoretic aspects of the developments of these fields, mainly on
the occurence of inverse cascades, i.e. generation of order from disorder. 
In this connection it is also discussed how the Silk effect (photon
diffusion) is counteracted by the inverse cascade, which moves energy from
smaller to larger scales.

Many galaxies (including our own) are observed to have magnetic fields.
One way to observe such fields is to study the polarization of light passing
the galaxy. Due to the interaction with the field and the plasma there
is a {\sl Faraday rotation} of the polarization vector, proportional to the 
field and to the square of the wave length of the light. In this way fields 
are found to have the order of magnitude 10$^{-6}-{\rm 10}^{-8}$ Gauss on a 
scale of 100 kpc\footnote{A p(arse)c is an astronomical unit, which has the 
physical value 1 pc$\approx$ 3.26 light year.} If 
you have forgotten what a G(auss) is: the mean field on the sun is 
approximately one G. 

Usually the galactic magnetic field is explained by the
{\sl dynamo effect}: turbulence (e.g. differential rotation) in the galaxy
enhances the magnetic field exponentially up to some saturation value,
corresponding to equipartition between kinetic and magnetic energy.
The dynamics which governs these phenomena is called 
{\sl magnetohydrodynamics}, abbreviated as MHD, which is essentially the 
Maxwell plus Navier-Stokes equations. The dynamo can produce an enhancement 
factor of several orders of magnitude.
An important feature is that the dynamo needs a {\sl seed} field. It appears
reasonable to assume that this field is of {\sl primordial origin}, i.e. it
has existed already in the early universe. Astrophysicists often say as a joke
that a primordial magnetic field is a field which has existed for so long that
everybody has forgotten how it was created. However, in particle physics
we must be more serious since we have knowledge of the early universe, and 
hence we should explain the origin of these fields.

\section{PRIMORDIAL MAGNETIC FIELDS IN THE EARLY UNIVERSE}
 
In natural units magnetic fields have dimension (mass)$^2$. At the 
electroweak scale, assuming the Higgs mass to be of order $m_W$, there is
essentially only one mass, $m_W$, and we may therefore expect something like
\begin{equation}
B_{\rm EW}\lesssim m_W^2\approx 10^{24}~{\rm G}
\end{equation}
on a scale $\sim 1/m_w$. This is a huge field, far larger than anything
one has ever seen or produced on this earth. How does this compare with the
rather weak fields found in galaxies?

In the standard cosmological model all distances are blown up by the scale
factor $R(t)$. It is useful for estimates that the scale factor is 
proportional to the inverse temperature. Thus, $R_{\rm now}/R_{\rm EW}=
T_{\rm EW}/T_{\rm now}\approx {\rm 10}^{15}$. Hence, an initial correlation
length of order $\sim 1/m_W$ is of order 1 cm today, which has no 
astrophysical interest. We need fields on a scale of order 100 kpc $\approx$
3$\times$10$^{23}$cm.

If we assume that $B$ is essentially random, we can estimate the field at any
distance from a simple random walk. We have the field at the initial 
correlation length, but we want it at $\approx$ 10$^{23}$ times this length.
Thus, in $d$ dimensions we have
\begin{equation}
<B_{\rm EW}>_{{\rm scale~10}^{23}/m_W}\approx 10^{24}~{\rm G}/({\rm 10}^{23})^{d/2}.
\end{equation}
So for $d=3$ we get $<B_{\rm EW}>\approx 10^{-10}$G, whereas for $d=2$ and 
$d=1$ we have $<B_{\rm EW}>$ approximately equal 10 G and 10$^{12}$G, 
respectively, on the scale of 10$^{23}/m_W$.

In order to see if these fields are reasonable, we need to know the 
cosmological developments of $<B>$. From MHD (with viscosity
ignored) one has the result that the flux through a surface
bounded by a curve following the fluid of charged particles is conserved.
Since such a surface increases like $R^2$, it follows that $<B>$ decreases
like $1/R^2$ when the universe expands with the scale factor $R$
\footnote{The metric is
\begin{equation}
d\tau^2=dt^2-R(t)^2\left[\frac{dr^2}{1-kr^2}+r^2d\Omega^2\right],
\end{equation}
where $k=+1,0,-1$ for a closed, flat or open universe, respectively.}.
It therefore follows that if today we need e.g. a primordial field of order
10$^{-15}$G on a scale of 100 kpc, then on the corresponding scale 
10$^{23}/m_W$ at the electroweak phase transition, we need $<B>\approx
10^{15}$ G. Thus, from the random walk estimates above we see that only
the case $d=1$ comes near this value, although a factor $10^{3}$ is
missing. Actually one could argue that the case $d=1$ is relevant, because 
in observing the magnetic field by Faraday rotation, a one dimensional
average is made along the line of sight. However, this argument is not really
convincing, since the dynamo effect is three dimensional, and hence the
field relevant for this effect is the very small 3$d$ average.

The conclusion is thus that if fields of the order of $m_W^2$ can be generated
at the EW-scale, then there could still be missing a factor of order $10^x$, 
where $x$ is of order 3. However, it should be emphasized that a field
of order $m_W^2$ is very large, and is not obtained in most mechanisms
for creation of primordial fields. Hence, in most cases $x$ is larger than 3. 

\section{MAGNETIC FIELDS FROM PARTICLE PHYSICS MECHANISMS}  

In this section we very briefly discuss a number of proposed mechanisms for 
the generation of primordial fields. The list is by no means exhaustive.  

\subsection{Fields From Inflation}

An inflationary creation of primordial magnetic fields has the advantage that
the coherence scale is larger than in other mechanisms. As an example,
we mention the work by Gasperini, Giovannini and Veneziano \refnote{1},
which is based on a pre-big-bang cosmology inspired by superstrings,
which is an alternative to the usual slow roll inflation. The dilaton
field $\phi$ in the Lagrangian
\begin{equation}
{\cal L}=-\sqrt{g}~e^{-\phi}(R+\frac{1}{2}~\partial_\mu \phi ~\partial^\mu\phi+
\frac{1}{4} F_{\mu\nu}F^{\mu\nu})
\end{equation}
amplifies the quantum fluctuations of $F_{\mu\nu}$. The magnetic energy
spectrum behaves like $\sim k^{0.8}$. The resulting magnetic fields
are of the right order of magnitude on a 100 kpc scale. Recently, however, 
Turner and Weinberg \refnote{2} have argued that this scenario requires fine 
tuning of the initial conditions in order to get enough inflation to solve
the flatness and horizon problems.

\subsection{Bubble Formation at the EW Phase Transition and Magnetic Fields}

In a first order EW phase transition bubbles of new vacuum are formed. This
was used by Baym, B\"odecker and McLerran \refnote{3} to obtain the generation
of a magnetic field. The main point is that the bubbles, although overall
neutral, have a dipole charge layer on the surface, so rotating bubbles 
generate a field.
Although the field from each bubble is very small, there is a large number of
bubbles, so depending on the subsequent development of $<B>$, in the end a 
reasonable magnitude can be produced. A different mechanism was considered by 
Kibble and Vilenkin \refnote{4}: when the bubbles collide, the phase of the 
Higgs field varies, giving rise to currents and a magnetic field. Again, in
this case one can get a reasonable magnitude provided the subsequent
development of $<B>$ is favourable.

\subsection{Superconducting Cosmic Strings and Other Mechanisms}

There exists a number of other mechanisms for the generation of magnetic 
fields. Vachaspati pointed out that if the gradients of Higgs fields
fluctuate, they can induce a magnetic field\refnote{5} at the electroweak
scale. The statistical 
averaging involved in this scenario was discussed in details by Enqvist and
me\refnote{6}. As mentioned already in the first section, the 
conclusion is that if line averaging is relevant, one obtains nearly the right
order of magnitude, since by this mechanism the field at genesis is of order
$m_W^2$ on a scale of $1/m_W$. However, this scenario operates with physically 
motivated fluctuations in gauge dependent quantities like gradients of Higgs 
fields. Such a procedure is not very clear to me. 

Recently there has been discussions of generation of primordial magnetic fields
from a network of Witten's superconducting cosmic strings 
\refnote{7}$^,$\refnote{8}. These
strings are current carrying, and hence produce magnetic fields. It turns
out that if the strings are created at the GUT phase transition, where the
current is very large, they can produce a field which is large enough over a 
sufficient scale, assuming that MHD does not give rise to any trouble. On the 
other hand, superconducting strings created at the EW phase transition cannot 
generate sufficient fields.

It has also been proposed that a Savvidy-type vacuum, where the energy is 
lowered relative to the trivial vacuum by having a magnetic field , can 
generate enough field\refnote{9}. For SU($N$) the field produced at a 
temperature $T$ is of order
\begin{equation}
B\sim T^2~\exp\left(-\frac{48\pi^2}{11Ng^2}\right).
\end{equation}
At the EW transition ($N=3$), this field is far too small. At the GUT 
transition, however, it produces a large enough field for $N=5$, due to the 
strong sensitivity of the exponent with respect to $N$. Whether this field is 
acceptable depends on the subsequent development according to MHD.

It was proposed long ago by Harrison that magnetic 
fields could be generated from vorticity present in eddies of plasma  
in the early universe\refnote{10}. This idea was criticized, and the eddies
were replaced by irrotational density fluctuations by Rees\refnote{10}. A more 
modern version of this scenario is due to Vachaspati and Vilenkin\refnote{10},
where the magnetic field is generated by vorticity arizing in the wakes of 
ordinary (i.e. not superconducting) cosmic strings.

Finally we mention that recently Joyce and Shaposhnikov\refnote{11} have 
presented a scenario which has the potential of leading to quite large fields.
The standard model has charges with abelian anomaly only (e.g. right-handed
electron number) which are essentially conserved in the very early
universe, until a short time before the EW transistion. A state with finite
chemical potential of such a charge is unstable to the generation of
hypercharge U(1) fields. Such fields can turn into large magnetic fields,
depending on their subsequent development.  

It is clear that the physical validity of most, if not all, of these scenarios,
depends on the subsequent non-linear development of the primordial field, due
to MHD. This will be discussed in the next section. In the end of this talk
we shall also discuss Silk diffusion, which is a mechanism for destroying
magnetic fields by turning it into heat. We shall show that this
linear diffusion is in fact counteracted by the non-linear terms of MHD.  

\section{INVERSE CASCADE FROM MAGNETOHYDRODYNAMICS}

We shall now investigate what happens subsequently to a primordial magnetic 
field generated in the early universe. To simplify things, here we give the
details only for the non-relativistic case, and mention without giving the
arguments, what happens in the general relativistic case.

\subsection{The Non-Relativistic MHD Equations}

In the rest frame of a plasma
consisting of charged particles with current ${\bf j}$, we have Ohm's law,
\begin{equation}
{\bf j}_{\rm rest}=\sigma~{\bf E}_{\rm rest},
\end{equation}
where $\sigma$ is the conductivity. The universe is a good conductor, so 
$\sigma$ is very large\footnote{In the relativistic era this can be seen from 
the following estimate: The current is defined by ${\bf j}=ne{\bf v}$. The 
velocity is given essentially by the Newtonian expression ${\bf v}/\tau
\approx e{\bf E}/E$, where $E$ is the relativistic energy, and $\tau$ is the
average time between collisions. Thus, $\tau\approx 1/n\sigma_{\times}$,
where $\sigma_{\times}$ is a typical relativistic cross section. Thus,
${\bf j}\approx ne(\tau e{\bf E}/E)$, so $\sigma\approx ne^2\tau/E\approx
e^2/E\sigma_{\times}$. A relativistic cross section goes like $e^4/T^2$,
since the temperature $T$ is a typical momentum transfer. Also, $E\sim
T$. Thus $\sigma\approx T/e^2$, which is very large in the early universe,
because the temperature is very high. At later stages the universe is still a 
good conductor, for different reasons \refnote{12}.}. Thus it follows that
\begin{equation}
{\bf E}_{\rm rest}\approx 0.
\end{equation}
In a frame moving with bulk velocity $\bf v$ one therefore has
\begin{equation}
{\bf E}\approx-{\bf v\times B}.
\end{equation}  
The induction equation $\partial {\bf B}/\partial t=-\nabla\times\bf E$
therefore gives
\begin{equation}
\partial{\bf B}/\partial t\approx \nabla\times(\bf v\times\bf B).~~(\rm MHD~I)
\end{equation}
This is one (out of two) of the fundamental MHD equations. It tells us that
the magnetic field is influenced by the velocity, and also, if you
start from $\bf B=0$ no magnetic field can be generated. Therefore
a seed field is needed in the dynamo mechanism.

The second fundamental MHD equation is the Navier-Stokes equation with the
Lorentz force $\bf j\times B$ on the right hand side. Here ${\bf j}$ can be 
estimated
from the Maxwell equation ${\bf j}+\partial {\bf E}/\partial t=\nabla\times
{\bf B}$. The time derivative can be estimated to be small in the 
non-relativistic case\footnote{We have $E\sim vB$, so $\partial E/\partial t
\sim (v/l)vB\sim (B/l)v^2$. But $|\nabla\times {\bf B}|\sim B/l$, where $l$ is
some typical length, so
the time derivative of the electric field can be ignored relative to the
curl of $B$.}, so ${\bf j}\approx \nabla\times{\bf B}$, so the Navier-Stokes
equation with the Lorentz force is given by
\begin{equation}
\partial \rho{\bf v}/\partial t +({\bf v\nabla})\rho{\bf v}\approx
-\nabla (p+\frac{1}{2}B^2)+({\bf B\nabla}){\bf B}.~~(\rm MHD~II)~~
\end{equation}
Here for simplicity we have ignored the viscosity. 
For $\sigma$ large, this can be generalized to the relativistic case at the
expense of a considerable increase in the complexity of the equations. 

\subsection{Why and When does MHD have an Inverse Cascade: A Simple Scaling
Argument}

Now let us suppose that by some particle physics mechanism a primordial
magnetic field is generated. At the genesis the field has some correlation
length, and the crucial question is then what happens as time passes. If the
correlation length grows smaller, corresponding to a {\sl cascade}, the
situation is quite bad, even for the inflationary scenario. Such a cascade
would appear if the system develops into a more chaotic direction. If, on the
other hand, we have an {\sl inverse cascade}, the correlation length grows
and the system develops towards more order. In an inverse cascade, energy is
thus transferred from smaller to larger scales.

It turns out that \refnote{13} the situation depends on the initial spectrum.
Roughly speaking, if the spectrum is concentrated at short (large) distances, 
it will develop into large (short) distances. To see this, one can make use of
the fact that the MHD equations are invariant under the ``self-similarity''
equations,
\begin{equation}
{\bf x}\rightarrow l{\bf x},~ t\rightarrow l^{1-h}t,~{\bf v}\rightarrow l^h
{\bf v},~{\bf B}\rightarrow l^h{\bf B},
\end{equation}
and if the viscosity $\nu$ and Ohmic resistance are included, we further have
\begin{equation}
\nu\rightarrow l^{1+h}\nu, ~\sigma\rightarrow l^{-1-h}\sigma.
\end{equation} 
From this it is very easy to show that the magnetic and kinetic energy
densities ($\cal E$), given in 3+1 dimensions by expressions like
\begin{equation}
{\cal E}(k,t)~V=\frac{2\pi k^2}{(2\pi)^3}\int d^3x~d^3y~
e^{i\bf k(x-y)}<{\bf B}({\bf x},t){\bf B}({\bf y},t)>,~{\rm with}~V=\int d^3x,
\end{equation}
where
\begin{equation}
V\int dk ~{\cal E}(k,t)=\frac{1}{2}\int d^3x <{\bf B}^2>={\rm total~
magnetic~energy},
\end{equation}
must satisfy the scaling relation\refnote{13}
\begin{equation}
{\cal E}(k/l,l^{1-h}t)=l^{1+2h}~{\cal E}(k,t).
\end{equation}
This is valid in the {\sl inertial range}, where viscosity and Ohmic 
resistance can be ignored. The general solution of this equation is
\begin{equation}
{\cal E}(k,t)=k^p~\psi(k^{(3+p)/2}t),
\label{scaling}
\end{equation}
with $p=-1-2h$ and $\psi$ some function of the single argument $k^{(3+p)/2}t$.
The interpretation of this equation is that, if at the initial time $t=0$ 
the spectrum is $k^p$ (from some particle physics mechanism), then at later 
times it will be governed by the function $\psi$. Hence the wave vector scales
like
\begin{equation}
k\sim t^{-2/(3+p)}
\end{equation}
Thus, if $p>-3$ there is an {\sl inverse cascade}, because $k$ moves towards
smaller values, whereas for $p<-3$, there
is a cascade. Thus, if initially we have a random system corresponding to
$p\geq 0$, then later the system becomes more ordered, as already announced.
For $p=-3$ it follows from eq. (15) that the $k$ and $t$ dependence of the
energy density become uncorrelated.

In the case when general relativity is included one obtains for a flat,
expanding universe\refnote{13}
\begin{equation}
R(t)^4~{\cal E}(k,t)=k^p~\psi(k^{(3+p)/2}~\tilde{t}),
\end{equation}
where $t$ is the Hubble time, $\tilde{t}=\int dt/R(t)$ is the conformal time, 
$\tilde{t}\propto\sqrt{t}$, and where $k$ is the {\sl comoving} wave vector, 
so that the physical wave vector is $k_{\rm phys}=k/R(t)$. Therefore the 
physical wave vector scales like
\begin{equation}
k_{\rm phys}\sim \tilde{t}^{-2/(3+p)}/R(t)\propto t^{-(5+p)/2(3+p)}.
\end{equation} 
Thus, if the spectrum starts out with $p=2$, corresponding to a Gaussian
random initial field\footnote{If the initial field is given by
\begin{equation}
<B_i({\bf x},0)B_k({\bf y},0)>=\lambda\left(\delta_{ik}-\frac{\partial_i
\partial_k}{\partial^2}\right)\delta^3({\bf x}-{\bf y}),
\end{equation}
then the initial energy is given by
\begin{equation}
{\cal E}(k,0)=\lambda (\delta_{ii}-k_ik_i/k^2)k^2=2\lambda k^2.
\end{equation}
Thus the general relativistic scaling goes as $k_{\rm phys}\sim t^{-3/5}$.},
we have a scaling of the physical wave vector by
$t^{-0.7}$, instead of $t^{-0.5}$ from pure expansion. If $p$ decreases,
the effective expansion increases. Thus, if the initial spectrum is 
characterized by $p=1$, we get a scaling $k_{\rm phys}\sim t^{-0.75}$, and
for $p=0$ we have $k_{\rm phys}\sim t^{-0.8}$. These examples imply 
physically that the typical size of an eddy increases like $t^{0.2}, 
t^{0.25}~{\rm and}~ t^{0.3}$, respectively, on top of the expansion factor.
Finally we mention that the ``scale invariant'' initial spectrum with $p=-1$
($dk/k=$scale inv.) has an increase of the typical eddy size by $t$, which
means that the eddies follow the horizon.

\subsection{Numerical Simulations in 2+1 Dimensions}

From the general scaling arguments presented above one cannot deduce the value
of the scaling function $\psi$. Here numerical investigations are needed, since
realistic analytic solutions of MHD are not known. However, a problem
arises, since the Reynold number\footnote{This number can be understood as
the ratio between ``typical'' non-linear terms and the linear viscosity term.
Thus, if $Re$ is large, turbulence is important.} is very large in the early 
universe. For example, in the paper by Brandenburg, Enqvist and me\refnote{14}
the magnetic Reynold number was estimated to be of order
10$^{17}$. In numerical simulations one cannot reach this value, no matter
how much computertime is used. We therefore did numerical simulations in
2+1 dimensions with an unrealistically low Reynold number\refnote{14}, taken to
be 10. So the non-linear terms are approximately ten times as important as
the diffusion terms. These terms are, however, needed to achieve numerical
stability (they act as a short distance cutoff).

We used the general relativistic MHD equations, which are considerably more
complicated than the MHD equations discussed in a previous section.
We took  the initial conditions
that the velocity vanishes and ${\bf B}$ is Gaussian random, so the
magnetic energy spectrum goes like $k$. Also, we took the energy
density to be $\rho={\rm const}/R^4$, and the pressure $p=\rho/3$. 

In fig. 1 the numerical results\refnote{14} are displayed. At the initial time
\begin{figure}[t]
\centering
\epsfxsize=9.2cm\epsfbox{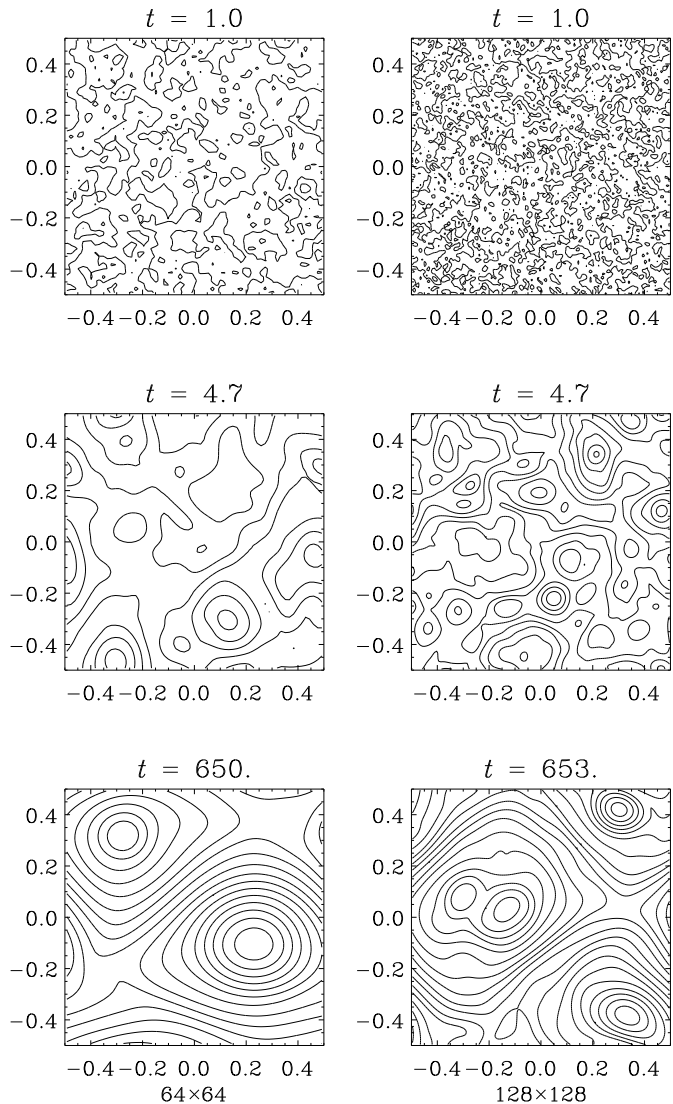}
\caption{Left column: magnetic field lines at different times at low 
resolution ($64\times64$ meshpoints).
Right column: magnetic field lines at different times at higher resolution
($128\times128$ meshpoints). This figure is taken from ref. 14.}
\end{figure}
there is a rather chaotic state, where the magnetic flux lines are either long
random walk curves, or small closed loops. We used periodic boundary
conditions, and satisfied div${\bf B}=0$. We see that in a short time the 
typical lenght scale increases considerably. In the end of the simulation 
there are quite large eddies. Therefore we clearly see an inverse cascade,
where order is produced from chaos, in contrast to the usual paradigm.

Also, the initial velocity ${\bf v}=0$ acquires a spectrum which shows an
inverse cascade. The velocity is initially induced by the Navier-Stokes
equation through the Lorentz force. The velocity generated this way then
influences the magnetic field through the induction equation, etc. etc.

\subsection{Numerical Simulations in 3+1 Dimensions: The Shell Model}

As already mentioned, simulations of MHD with large Reynold numbers is not
possible with present day computers\footnote{This also applies to
hydrodynamics, and is perhaps the reason why weather forecasts are pretty bad,
at least in Denmark.}. The situation gets worse when we go from 2+1
to 3+1 dimensions. Therefore one needs to make a model which has as many
features of the real Navier-Stokes as is compatible with practical
tractability. In recent years the so-called GOY (Gledzer, Ohkitani and
Yamada) model has become increasingly popular. Another name for this
model is the ``shell model''. It gives results in good agreement
with experiments, especially as far as the subtle intermittency effects are
concerned. The model captures a basic feature of turbulence, namely the
coupling of many different length scales. It is not known whether the model
has relation to the real Navier-Stokes and MHD. But it nicely illustrates the
behaviour of a system in which numerical simulations are made 
difficult by the effect of a huge number of couplings between the different
length scales. Also, real world conservation laws (energy, helicity) are 
buildt into the model. 

To motivate the model, let us mention that in the Navier-Stokes equations and
MHD one has terms like
\begin{equation}
({\bf v\nabla)v},~~{\bf \nabla\times(v\times B)},~~({\bf B\nabla)B},
\end{equation}
etc. In Fourier space they e.g. have the form
\begin{equation}
({\bf v\nabla)v}\rightarrow \int d^3p~v_i({\bf p})(p_i-k_i) v_j({\bf p-k}).
\end{equation}
Experience with numerical simulations show that the largest contributions
come from triangles in ${\bf k}-$space with similar side lengths. This is
taken as a ``phenomenological'' input in the shell model.

At this stage, for numerical purposes, one would discretize ${\bf k}-$space.
In the shell model one of the basic ingredients is a hierachical structure,
where $|{\bf k}|-$space is divided into shells
\begin{equation}
k_n=\lambda^nk_0,~~n=1,2,...,N.
\end{equation}
Here $\lambda$ is often taken to be 2. There furthermore exists a complex
``velocity mode'' $v_n=v(k_n)$, which can be considered as the Fourier
transform of the velocity difference $|{\bf v}(|{\bf x}|+2\pi/\lambda)-
{\bf v}(|{\bf x}|)|$.
Since $k_n$ increases exponentially, it covers a wide range of corresponding
length scales. The model then assumes couplings between neighbours and
next nearest neighbours,
\begin{equation}
({\bf v\nabla)v}\rightarrow \sum_{i,j=-2}^{i,j=2}c_{ij}~v_{n+i}k_nv_{n+j},
\end{equation}
where the sum is over neighbours and/or next nearest neighbours to $n$. The 
couplings $C_{ij}$ in this sum should be made such that energy is conserved 
in the absence of diffusion. Thus, energy conservation
\begin{equation}
\int (v^2+B^2)d^3x={\rm const}
\end{equation}
now corresponds to
\begin{equation}
\sum_{n=1}^{n=N}(|v_n|^2+|B_n|^2)={\rm const}.
\end{equation}
Thus, we need to satisfy
\begin{equation}
\sum_{n=1}^{n=N}\left(v_n\frac{dv_n^*}{d\tilde{t}}+B_n\frac{dB_n^*}
{d\tilde{t}}+{\rm complex~conj.}\right)=0,
\end{equation}
where, as before, $\tilde{t}$ is the conformal time, $\tilde{t}=\int dt/R(t)$.
In this approach the vectorial character is thus lost, but the conservation of
energy is kept as an essential feature. 

We should now find equations for the time derivatives respecting the
conservation of energy. Taking into account some factors from general
\begin{figure}[t]
\centering
\epsfxsize=9cm\epsfbox{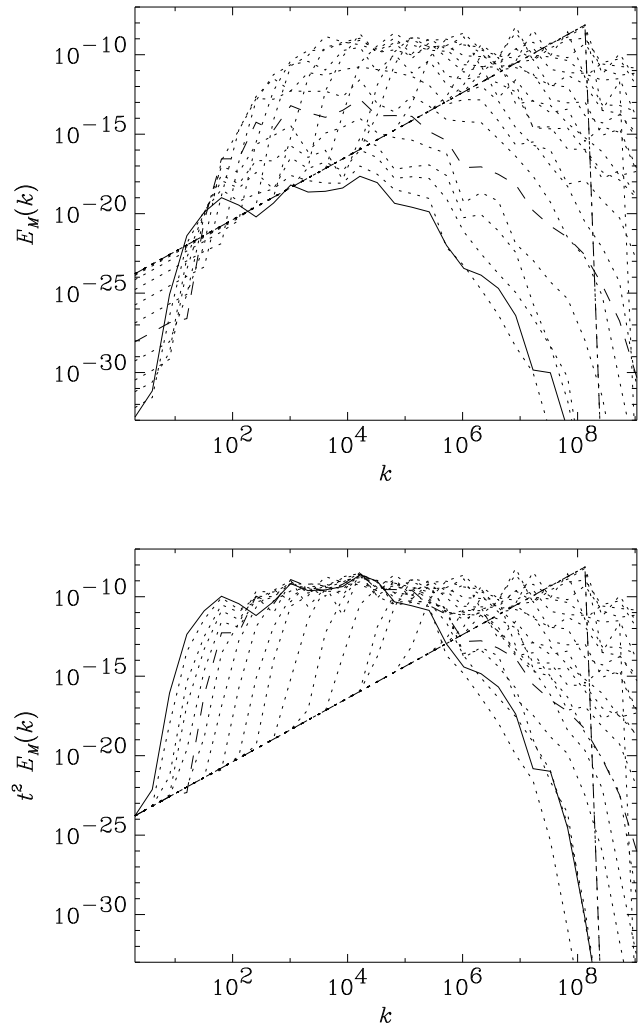}
\caption{
Spectra of the magnetic energy at different times.
The straight dotted-dashed line gives the initial condition ($t_0=1$),
the solid line gives the final time ($t=3\times10^4$), and the dotted
curves are for intermediate times (in uniform intervals of 
$\Delta\log(t-t_0)=0.6)$. $A=1$, $B=-1/2$, and $C=0$. This figure is from ref.
14.}
\end{figure}
relativity in an expanding universe (the expansion factor as well
as the energy density and pressure) we get\refnote{14}
\begin{equation}
\begin{array}{lll}
\frac{8}{3}dv_n/d\tilde{t}&=ik_n(A+C)
(v^*_{n+1}v^*_{n+2}-B^*_{n+1}B^*_{n+2})\\
                        &\!\!\!+ik_n(B-{\textstyle{1\over2}}C)
(v^*_{n-1}v^*_{n+1}-B^*_{n-1}B^*_{n+1})\\
                        &\!\!\!\!\!\!\!-ik_n({\textstyle{1\over2}}B+
{\textstyle{1\over4}}A)
(v^*_{n-2}v^*_{n-1}-B^*_{n-2}B^*_{n-1}),
\end{array}
\end{equation}
\begin{equation}
\begin{array}{lll}
dB_n/d\tilde{t}&=ik_n(A-C)
(v^*_{n+1}B^*_{n+2}-B^*_{n+1}v^*_{n+2})\\
                        &\!\!\!+ik_n(B+{\textstyle{1\over2}}C)
(v^*_{n-1}B^*_{n+1}-B^*_{n-1}v^*_{n+1})\\
                        &\!\!\!\!\!\!\!-ik_n({\textstyle{1\over2}}B
                                            -{\textstyle{1\over4}}A)
(v^*_{n-2}B^*_{n-1}-B^*_{n-2}v^*_{n-1}),
\end{array}
\end{equation}
where with $A,B,C$ arbitrary constants energy is conserved. In 3+1
dimensions, magnetic helicity is also conserved. In the continuum helicity is 
given by
\begin{equation}
H=\int d^3x ~{\bf AB},
\end{equation}
where {\bf A} is the vector potential. This conservation is trivial in 2+1
dimensions, since there $H=0$. To mimic conservation of $H$ in the shell model
we require that the quantity
\begin{equation}
H_{\rm shell}=\sum_{n=1}^{n=N} (-1)^nk_n^{-1}B^*_nB_n 
\end{equation}
is conserved. The reason is that $k_n^{-1}B_n$ is like the vector potential.
The factor $(-1)^n$ is a more ``phenomenological'' factor. The corresponding
conservation in hydrodynamics ($\int {\bf v(\nabla\times v)}d^3x=$const) has 
been studied, and it was found that the integrand oscillates in sign. This
is then taken into account in the shell model by the oscillating factor.

The requirement that helicity is conserved thus corresponds to taking into 
account 3+1 dimensions, and it leads to the following values
for the otherwise arbitrary constants $A,B,C$,
\begin{equation}
A=1,~~B=-1/4,~~C=0.
\end{equation}
Using these values, we have 2$N$ coupled set of equations. In our calculations
we took $N=$30, corresponding to solving 60 coupled equations. The resulting
spectra at different times are shown in fig. 2. Again we see a nice inverse   
\begin{figure}[tbp]
\centering
\epsfxsize=9cm\epsfbox{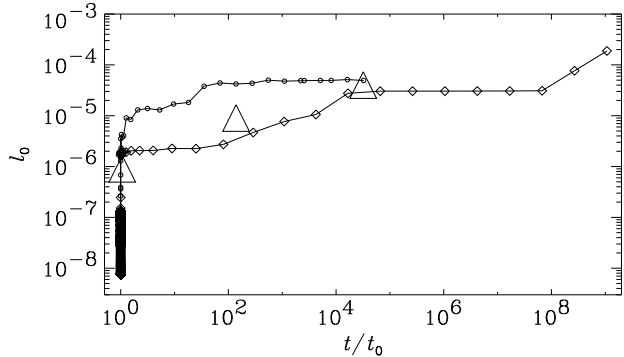}
\caption{The correlation length (the ``integral scale'') as a function of
time. The two curves are for slightly different models. This figure is from 
ref. 14.}
\end{figure}
cascade, because as functions of the comoving wave vector $k$ the spectra 
clearly move towards $k=0$.

To give a more precise picture of the change of the spectrum towards large 
distances, we also computed a correlation length defined by averaging over
the magnetic energy density,
\begin{equation}
l_0\equiv \int dk\frac{2\pi}{k}{\cal E}(k,t)\left[\int dk~{\cal E}(k,t)
\right]^{-1}.
\end{equation}
In turbulence theory this quantity is called the ``integral scale''. It is
a measure of the characteristic size of the largest eddies of turbulence.

The result is shown in fig. 3. We see that initially the system moves extremely
rapidly towards larger scales. Clearly MHD (in the shell version) does not
like the initial Gaussian random state for the magnetic field! The scaling
arguments in eq. (18) predicts an increase in the eddy size like $t^{0.2}$.
This cannot directly be compared to the integral scale $l_0$, since the
integrations in eq. (33) are limited by an ultraviolet cutoff, which also
becomes scaled. However, a fit in ref. 14 gives $l_0\sim t^{0.25}$, if
the steep initial increase in fig. 3 is ignored. Taking into account some
uncertainty in the fitting, this is in good agreement with the scaling in 
eq. (18).

\section{EFFECTS OF DIFFUSION: SILK DAMPING}

The effect of diffusion has been ignored in the above discussion, except as
a short distance cutoff in the numerical calculations. However, this is not
realistic, as was pointed out by Siegl, Olinto, and Jedamzik\refnote{15}. This
is connected to Silk damping, which occurs in the charged plasma because
radiation can penetrate the plasma and carry away momentum by scattering
off the charged particles. Around the time of recombination photon diffusion 
became very important and corresponded to a very large photon mean free path.
The diffusion coefficient is proportional to the photon mean free path, and 
hence photon diffusion at that time cannot be ignored\refnote{15}.  
In a linear approximation of MHD it was clearly demonstrated that the magnetic
field must be destroyed, the magnetic energy beeing turned into 
heat\refnote{15}. Silk diffusion would therefore remove the hope of 
understanding primordial magnetic fields from most points of view!

All hope is not lost, however, since the {\sl non-linear} inverse cascade, 
discussed in the previous section, {\sl counteracts Silk diffusion.} While the
latter is buzy removing magnetic energy at shorter scales, the
former is active in removing the energy from these scales to large scales. 
As we have seen in fig. 3, this happens very quickly. Therefore,
without doing any calculations it is clear that these two mechanisms 
compete against one another.

To be more precise, numerical simulations are needed. This was done by
Brandenburg, Enqvist, and me\refnote{16}, and the result is that
even if Silk diffusion is included, the inverse cascade is strong enough
to make the magnetic field survive, at least until close to recombination.
This should be enough for the dynamo effect to start to operate. We refer
to the original paper\refnote{16} for a full discussion of this.

\section{CONCLUSIONS}

In conclusion we mention that there are several particle physics models
which can produce primordial magnetic fields. Of course, they are based on
assumptions which may not turn out to be ultimately true. For example, there
may not be a first order EW phase transition, superconducting or
ordinary cosmic strings may not exist, etc. etc. So when the dust settles,
there may not be so many mechanisms which survive. Also, it should be 
remembered that without the inverse cascade, there is no hope to produce
large enough background fields (this does, of course, not apply to 
the inflationary mechanism), and it may be that for some or all of these
models, the inverse cascade is not large enough. 

Thus, in estimating the effect of various models one should take
into account the combined effect of the inverse cascade and Silk
diffusion. This will perhaps require rather complicated numerical
calculations, although some results might conceivably be obtained or guessed
from simple scaling arguments, as discussed in ref. 13.  

Finally, we mention that there is a very interesting proposal for direct
observation of a primordial background field\refnote{17}. The idea is that
gamma rays arising from strong sources can scatter in a background field,
making pair production and delayed photons. The spectrum of these photons
could then be observed, provided the field is of order 10$^{-24}$ G or
larger\refnote{17}. If this is technically feasible, important information
on the spectrum would be obtained, which could then be compared with different
models.

\end{document}